\documentclass[12pt,number]{elsarticle}
\usepackage{graphicx}
\usepackage{amsmath}
\usepackage{amsfonts}
\usepackage{amssymb}
\usepackage{longtable}

\begin{document}

\journal{Phys. Lett. A}

\begin{frontmatter}
\title{Neutral and cationic free-space oxygen-silicon clusters SiO$_n$ ($1<n\leq6$), and possible relevance to crystals of SiO$_2$ under pressure}
\author[CTChem]{G. Forte}
\author[CTPhys,SSC,CNISM,INFN]{G. G. N. Angilella\corref{corr}}
\ead{giuseppe.angilella@ct.infn.it}
\author[CTChem]{V. Pittal\`a}
\author[AntwerpPhys,Oxford]{N. H. March}
\author[CTPhys,CNISM]{R. Pucci}

\address[CTChem]{Dipartimento di Scienze del Farmaco,\\
Facolt\`a di Farmacia, Universit\`a di Catania,\\
Viale A. Doria, 6, I-95126 Catania, Italy}
\address[CTPhys]{Dipartimento di Fisica e Astronomia, Universit\`a di Catania,\\
64, Via S. Sofia, I-95123 Catania, Italy}
\address[SSC]{Scuola Superiore di Catania, Universit\`a di Catania,\\ Via S.
Nullo, 5/i, I-95123 Catania, Italy}
\address[CNISM]{CNISM, UdR Catania, 64, Via S. Sofia, I-95123 Catania, Italy}
\address[INFN]{INFN, Sez. Catania, 64, Via S. Sofia, I-95123 Catania, Italy}
\address[AntwerpPhys]{Department of Physics, University of Antwerp,\\
Groenenborgerlaan 171, B-2020 Antwerp, Belgium}
\address[Oxford]{Oxford University, Oxford, UK}
\cortext[corr]{Corresponding author.}

\begin{abstract}
Motivated by the theoretical study of Saito and Ono (2011) on three crystalline
forms of SiO$_2$ under pressure, quantum-chemical calculations on various
free-space clusters of SiO$_n$ and GeO$_n$ for $1<n\leq6$ are reported here.
Both neutral and cationic clusters have been examined, for both geometry and
equilibrium bond lengths. Coupled clusters and correlation-corrected MP2
calculations are presented. For the cations, we emphasize especially the
structural distortions occurring in removing degeneracies.

\end{abstract}

\end{frontmatter}

\section{Introduction}
\label{sec:introduction}

The background to the present study is to be found in the theoretical work of
Stenhouse \emph{et al.} in this Journal \cite{Stenhouse:76}. This dealt with
scattering intensities and partial structure factors in vitreous silica, both
X-ray and neutron diffraction experiments being invoked. Our interest in the
present Letter focusses on the electron density distribution in free-space
clusters such as SiO$_2$, SiO$_4$, and SiO$_6$. This quantum-chemical study has
been directly motivated by the theoretical work of Saito and Ono \cite{Saito:11}
on different crystalline forms of SiO$_2$ under pressure (see also \cite{Tse:11}
on vitreous silica). In addition to our calculations on the above three neutral
clusters of the form SiO$_n$, we have also considered cationic free-space
clusters of this kind.

\section{Free-space cluster SiO$_2$ by Hartree-Fock (HF) theory}

All the clusters studied in this work have been first considered in free space,
by performing geometry optimization within the coupled cluster approximation
with single and double excitations (CCSD) \cite{Cizek:69,Purvis:82,Scuseria:88}.
Correlation-corrected MP2 calculations were then carried out for the
optimization. In all the calculations, performed by means of the Gaussian~09
software \cite{Frisch:09}, the 6-31 basis set, supplemented by polarization and
diffuse functions (6-31G+**), was used.

\begin{figure}[t]
\centering
\includegraphics[width=0.4\columnwidth]{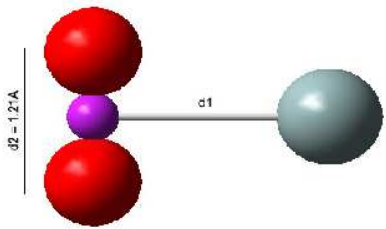}
\caption{(Color online) Shows Si atom approaching O$_2$ molecule.}
\label{fig:1}
\end{figure}

Fig.~\ref{fig:1} shows the way in which an O$_2$ molecule, at a fixed bond
length $1.21$~\AA, is brought up to a single Si atom in the neutral cluster
SiO$_2$. The distance $d$ was varied from 2~\AA{} to 1.75~\AA{} at intervals of
0.05~\AA. Table~\ref{tab:1} reports the HF energies for singlet, triplet, and
quintet spin states. While O$_2$ still retains its spin density at the largest
distance $d_1 = 2$~\AA, the triplet state energy lying below the singlet value,
a crossover in energy taking place around 1.85~\AA; at 1.75~\AA{} separation
the singlet state clearly lies lowest.

\begin{table}[t]
\centering
\begin{tabular}{cccc}
\hline 
$d_1$ (\AA) & \multicolumn{3}{c}{$E$ (Hartree)} \\
            & singlet & triplet & quintet \\
\hline 
$2.00$ & $-439.104$  &   $-439.118$  &   $-439.013$ \\
$1.95$ & $-439.112$  &   $-439.122$  &   $-439.002$ \\
$1.90$ & $-439.121$  &   $-439.125$  &   $-438.990$ \\
$1.85$ & $-439.129$  &   $-439.127$  &   $-438.976$ \\
$1.80$ & $-439.137$  &   $-439.128$  &   $-438.960$ \\
$1.75$ & $-439.144$  &   $-439.128$  &   $-438.941$ \\
\hline
\end{tabular}
\caption{Energies in Fig.~\ref{fig:1} as a function of distance $d_1$, for
different spin states.}
\label{tab:1}
\end{table}

\section{Neutral free-space clusters SiO$_4$}

\begin{figure}[t]
\centering
\includegraphics[width=0.4\columnwidth]{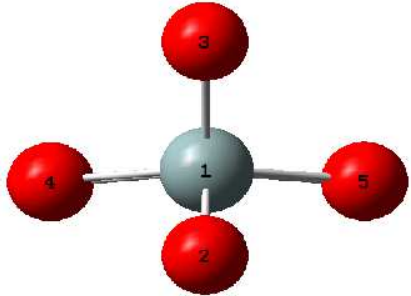}
\caption{(Color online) SiO$_4$ optimized low-lying isomer. For precise bond
lengths and angles, see Table~\ref{tab:2}.}
\label{fig:2}
\end{figure}

Fig.~\ref{fig:2}, calculated by HF theory, shows what we consider to be the
low-lying isomer of the free-space cluster SiO$_4$. The geometry of this present
proposal for the ground state is recorded in Table~\ref{tab:2}. It is seen that
there are two SiO distances which we determine to be 1.64~\AA{} and 1.73~\AA{}
respectively. The corresponding angles range from 82.3$^\circ$ to 130$^\circ$,
via the (almost) tetrahedral angle of 109$^\circ$. It is already relevant to
make some contact with the crystalline calculations of Saito and Ono
\cite{Saito:11} (see especially their Table~II). These workers show that at
ambient temperature and pressure, the ground-state structure $\alpha$-quartz
($q$-SiO$_2$ in their Table~II) has two slightly different bond lengths
1.60~\AA{} and 1.61~\AA, with angles near the tetrahedral value $\sim
110^\circ$. Above a pressure of 2~GPa, a rutile structure ($r$-SiO$_2$) is
formed, and Table~II records two SiO bond lengths of 1.75~\AA{} and 1.79~\AA,
which are at least fairly near to our predicted second bond length 1.73~\AA{}
for the free-space SiO$_4$ cluster in Fig.~\ref{fig:2}. Angles were not recorded
however for $r$-SiO$_2$ in their Table~II.

We shall return briefly to our further results pertaining to SiO$_4$ clusters when we summarize our calculations on cationic free-space
clusters below.

\begin{table}[t]
\centering
\begin{tabular}{cccc}
\hline
 & distance (\AA) & & angle ($^\circ$) \\
\hline
1--2	& 1.729	& 	  2--1--3	&   82.25 \\
1--3	& 1.729	& 	  2--1--4	&   108.5 \\
1--4	& 1.639	& 	  2--1--5	&   108.5 \\
1--5	& 1.639	& 	  4--1--5	&   130.14 \\
\hline
\end{tabular}
\caption{Bond lengths and angles for optimized low-lying isomer of SiO$_4$, as
shown in Fig.~\ref{fig:2}.}
\label{tab:2}
\end{table}

\begin{figure}[t]
\centering
\includegraphics[width=0.4\columnwidth]{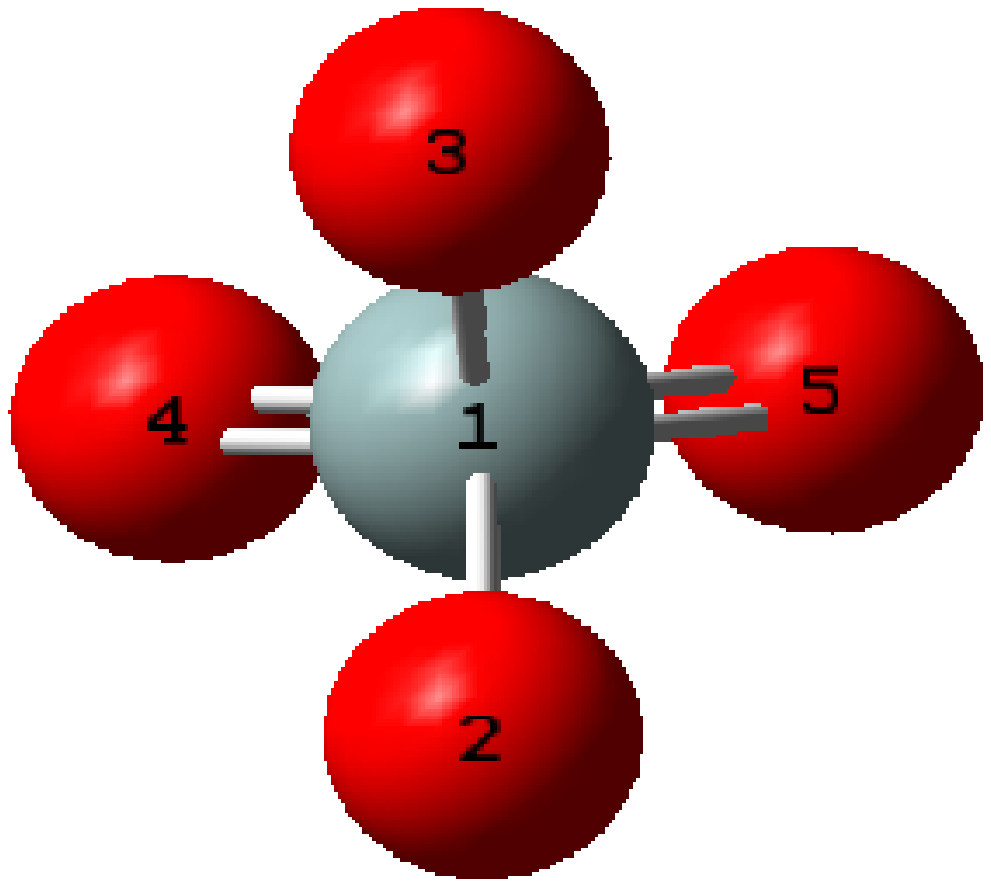}
\includegraphics[width=0.4\columnwidth]{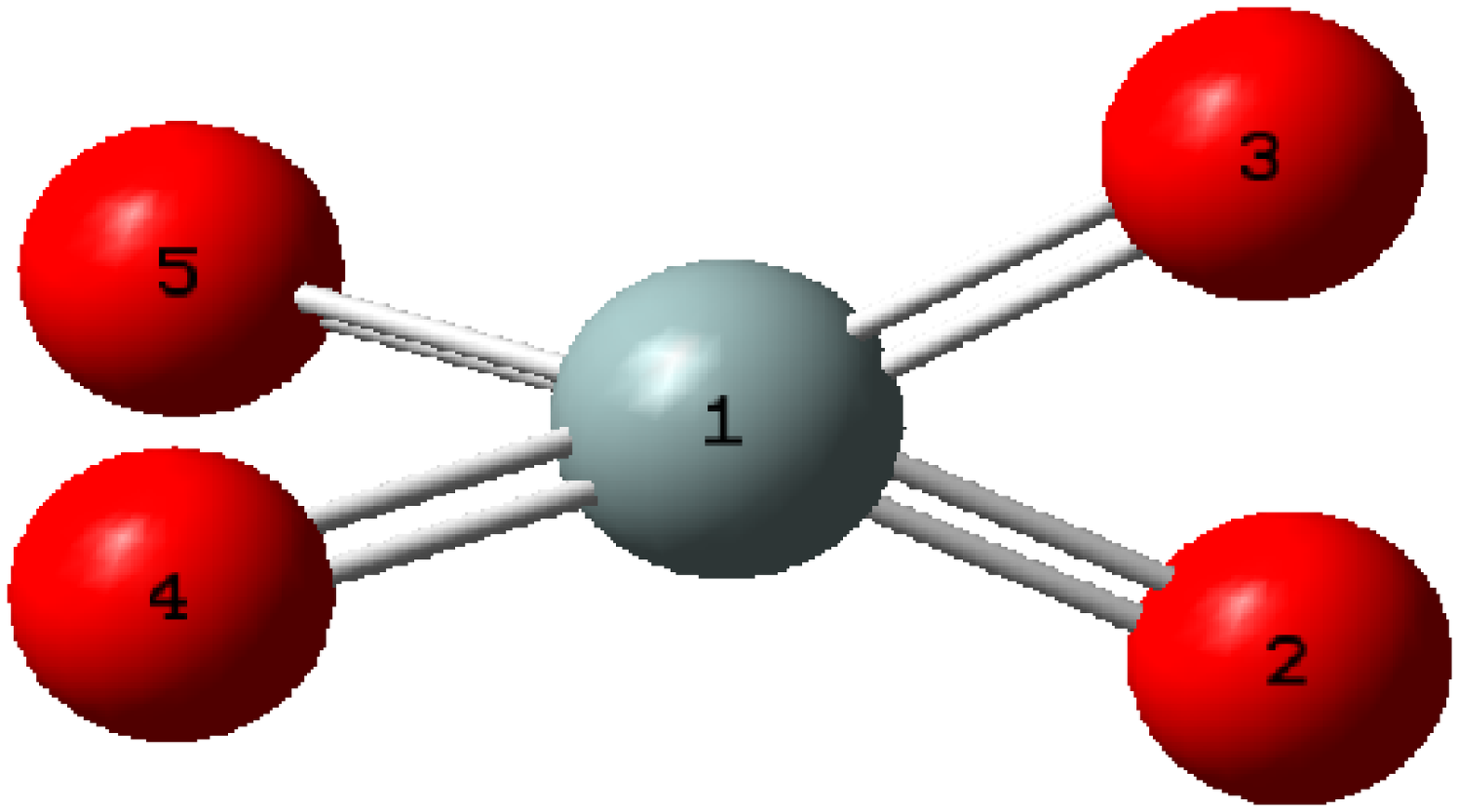}
\caption{(Color online) Shows two different views of isomer of SiO$_4$, with
geometry as in Table~\ref{tab:2}.}
\label{fig:3multi}
\end{figure}

\section{Neutral free space cluster of a Si atom with six surrounding O atoms}

\begin{figure}[t]
\centering
\includegraphics[width=0.4\columnwidth]{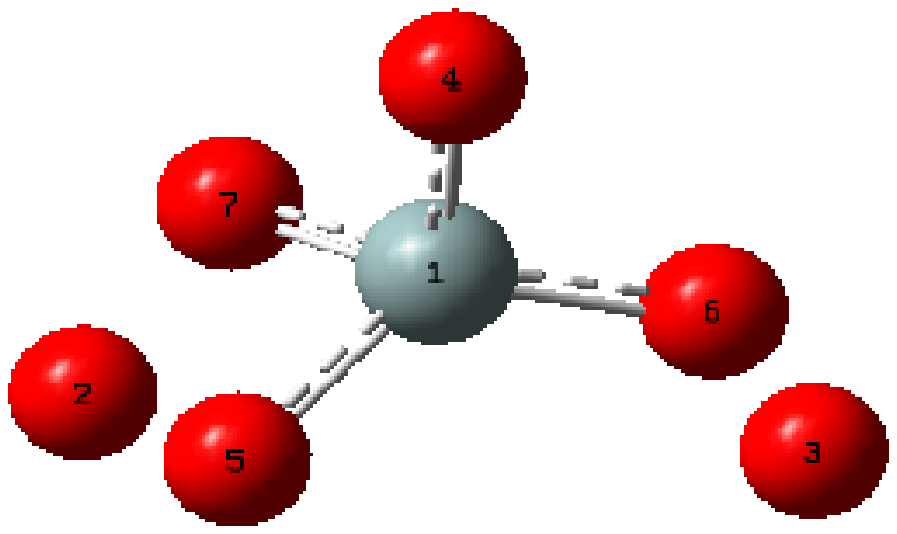}
\includegraphics[width=0.4\columnwidth]{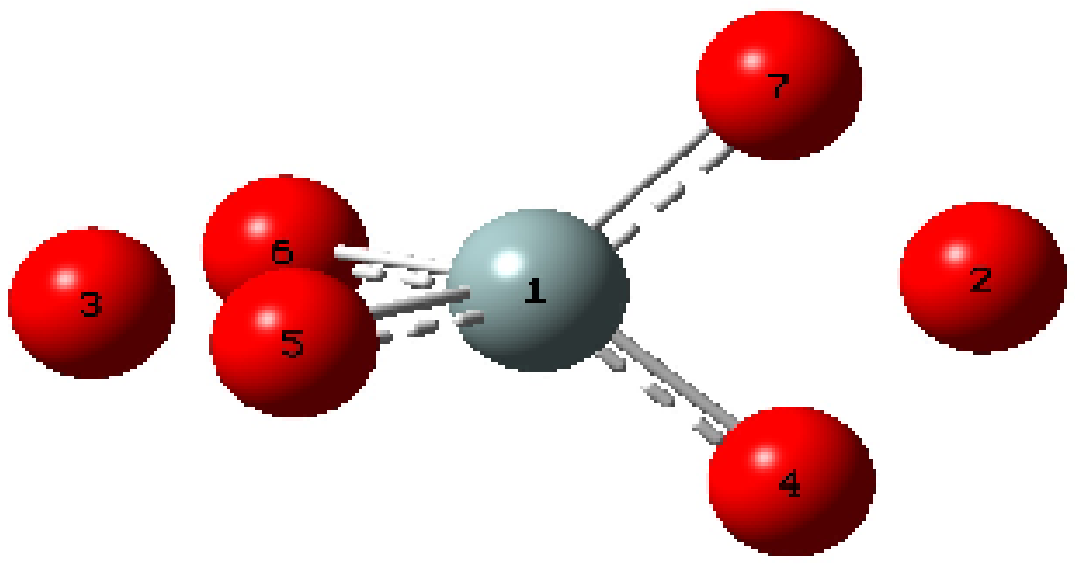}
\begin{minipage}[c]{0.45\columnwidth}
\begin{center}
(a)
\end{center}
\end{minipage}
\begin{minipage}[c]{0.45\columnwidth}
\begin{center}
(b)
\end{center}
\end{minipage}
\caption{(Color online) Proposed low-lying isomers of SiO$_6$.}
\label{fig:3}
\end{figure}

Fig.~\ref{fig:3}a shows first a neutral cluster in which atom 1 is Si and
tetrahedral angles for 4 surrounding O atoms are assumed. The energy and bond
distances are recorded in Table~\ref{tab:3}, the larger distance 1--2 and 1--3
being around $2.3\pm 0.1$~\AA, compared with 4 shorter bond lengths around
$1.66\pm 0.01$~\AA. Table~\ref{tab:4} shows how variation of the geometry away
from the tetrahedral angle $109.5^\circ$ entering Table~\ref{tab:3} lowers the HF
energy by $\sim 0.008$~Hartree, with the cluster form depicted now in
Fig.~\ref{fig:3}b. We note briefly that in Table~II of Ref.~\cite{Saito:11}, the
longest bond length recorded is for $r$-SiO$_2$, and is 1.81~\AA. We have also
considered the stability of the neutral cluster SiO$_6$ with respect to an
isolated Si atom plus two ozone molecules O$_3$, finding a relative energy
difference of $\Delta E=-0.373$~hartrees.

\begin{table}[t]
\centering
\begin{small}
\begin{tabular}{cccccc}
\hline
$n$ & $E(\mathrm{O})$ & $E(\mathrm{Si})$ & $E(\mathrm{Si} + n \mathrm{O})$ &
$E(\mathrm{SiO}_n )$ & $\Delta E$ \\
\hline
2 & $-74.940$ & $-289.017$ & $-438.897$ & $-439.244$ &
$-0.346$ \\
4 & $-74.940$ & $-289.017$ & $-588.777$ & $-589.384$ &
$-0.607$ \\
\hline
\end{tabular}
\end{small}
\caption{Stability with respect to the elements of isomers SiO$_2$ and SiO$_4$
displayed in Figs.~\ref{fig:1} and \ref{fig:2}, respectively. Energies are in
Hartrees.}
\label{tab:3}
\end{table}

\begin{table}[t]
\centering
\begin{tabular}{cccc}
\hline
 & distance (\AA) & & angle ($^\circ$) \\
\hline
\hline
\multicolumn{4}{c}{SiO$_6$ (Fig.~\ref{fig:3}a), $E= -739.4260$~Hartree} \\
\hline 
1--4	& 1.647 &     5--1--4	&   109.5 \\
1--5	& 1.669 & 	  5--1--6	&   109.5 \\
1--6	& 1.67  & 	  5--1--7	&   109.5 \\
1--7	& 1.679 & 	  	  	& \\
1--2	& 2.195 & 	  	  	& \\
1--3	& 2.377 & 	  	  	& \\
\hline
\multicolumn{4}{c}{SiO$_6$ (Fig.~\ref{fig:3}b), $E= -739.4343$~Hartree} \\
\hline
1--4  &   1.664   & 5--1--4	  &   123.07 \\
1--5  &   1.664   & 5--1--6	  &   84.77  \\
1--6  &   1.664   & 5--1--7	  &   123.07 \\
1--7  &   1.664   &       	  & \\
1--2  &   2.208   &       	  & \\
1--3  &   2.208   &       	  & \\
\hline
\end{tabular}
\caption{Shows geometries of low-lying isomers of SiO$_6$ depicted in
Fig.~\ref{fig:3}.}
\label{tab:4}
\end{table}

\section{Some less detailed results with the Si atom replaced by Ge in free-space neutral clusters}

\begin{figure}[t]
\centering
\includegraphics[width=0.4\columnwidth]{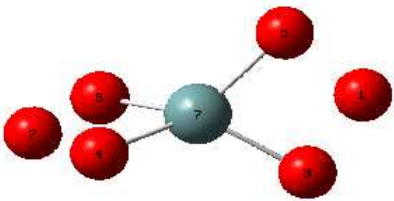}
\caption{(Color online) Low-lying isomer for GeO$_6$, with bond lengths and
angles as in Tab.~\ref{tab:5}.}
\label{fig:4}
\end{figure}

Prompted by the solid-state theoretical calculations presented by Saito and Ono \cite{Saito:11}, we have also studied GeO$_n$ clusters, though
in somewhat less detail than their Si counterparts. One of the configurations of the free-space neutral cluster is depicted in
Fig.~\ref{fig:4} for GeO$_6$. The bond distances and angles are collected in Table~\ref{tab:5}, for comparison with SiO$_6$ for the neutral
cluster shown in Fig.~\ref{fig:3}b.

\begin{table}[t]
\centering
\begin{tabular}{cccc}
\hline
 & distance (\AA) & & angle ($^\circ$) \\
\hline
1--7	& 2.333 & 3--7--4  & 	123.54 \\
2--7	& 2.333 & 3--7--5  & 	130.04 \\
3--7	& 1.797 & 3--7--6  & 	78.84 \\
4--7	& 1.797 & 	    	 &  \\
5--7	& 1.796 & 	    	 &  \\
6--7	& 1.797 & 	    	 &  \\
\hline
\end{tabular}
\caption{Bond lengths and angles for optimized low-lying isomer of GeO$_6$, as
shown in Fig.~\ref{fig:4}.}
\label{tab:5}
\end{table}

\section{Singly-charged positive ion clusters SiO$_4^+$ and GeO$_4^+$}

Having treated the neutral clusters of SiO$_n$ in some detail, we felt it of
interest in longer terms to estimate bond length distortions due to ionization
for both SiO$_4^+$ and GeO$_4^+$. We note first that if we were dealing with the
free-space tetrahedral molecule silane, then we could appeal to experiment for
the symmetry of the cation SiH$_4^+$ (see, \emph{e.g.,}
Ref.~\cite{Krishtal:11}).

\begin{figure}[t]
\centering
\includegraphics[width=0.4\columnwidth]{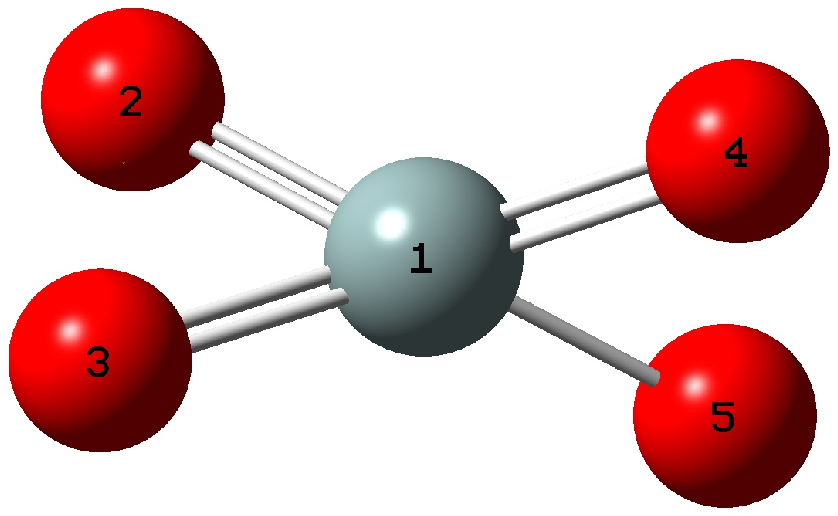}
\includegraphics[width=0.4\columnwidth]{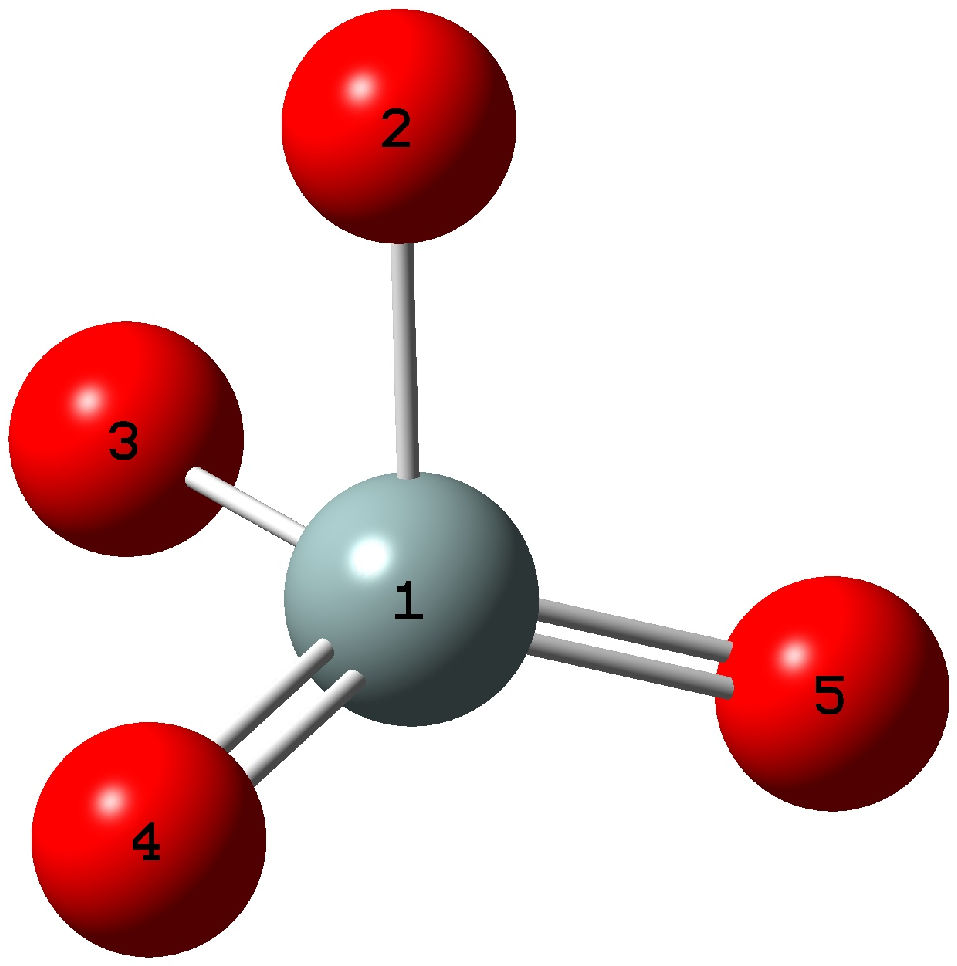}
\caption{(Color online) Low-lying isomers predicted theoretically for neutral
SiO$_4$ and cationic SiO$_4^+$, the total ground-state energies being
$E=-589.3844$ and $E=-588.8507$~a.u., respectively. Bond lengths and angles are
given in Tab.~\ref{tab:SiO4p}.}
\label{fig:SiO4p}
\end{figure}

\begin{table}[t]
\centering
\begin{tabular}{cccc}
\hline
 & distance (\AA) & & angle ($^\circ$) \\
\hline
\multicolumn{4}{l}{SiO$_4$} \\
1--2	&1.63  &2--1--4	&  137.12\\
1--3	&1.63  &2--1--5	&  137.13\\
1--4	&1.63  &2--1--3	&  62.24 \\
1--5	&1.63  &	  	&	    \\
\hline
\multicolumn{4}{l}{SiO$_4^+$} \\
1--2 &  1.729 & 2--1--4  &	108.5 \\
1--3 &  1.729 & 2--1--5  &	108.52\\
1--4 &  1.639 & 2--1--3  &	82.25 \\
1--5 &  1.639 &	    	 &		 \\
\hline
\end{tabular}
\caption{Bond lengths and angles for neutral cluster SiO$_4$ and cation
SiO$_4^+$, as shown in Fig.~\ref{fig:SiO4p}.}
\label{tab:SiO4p}
\end{table}

\begin{figure}[t]
\centering
\includegraphics[width=0.4\columnwidth]{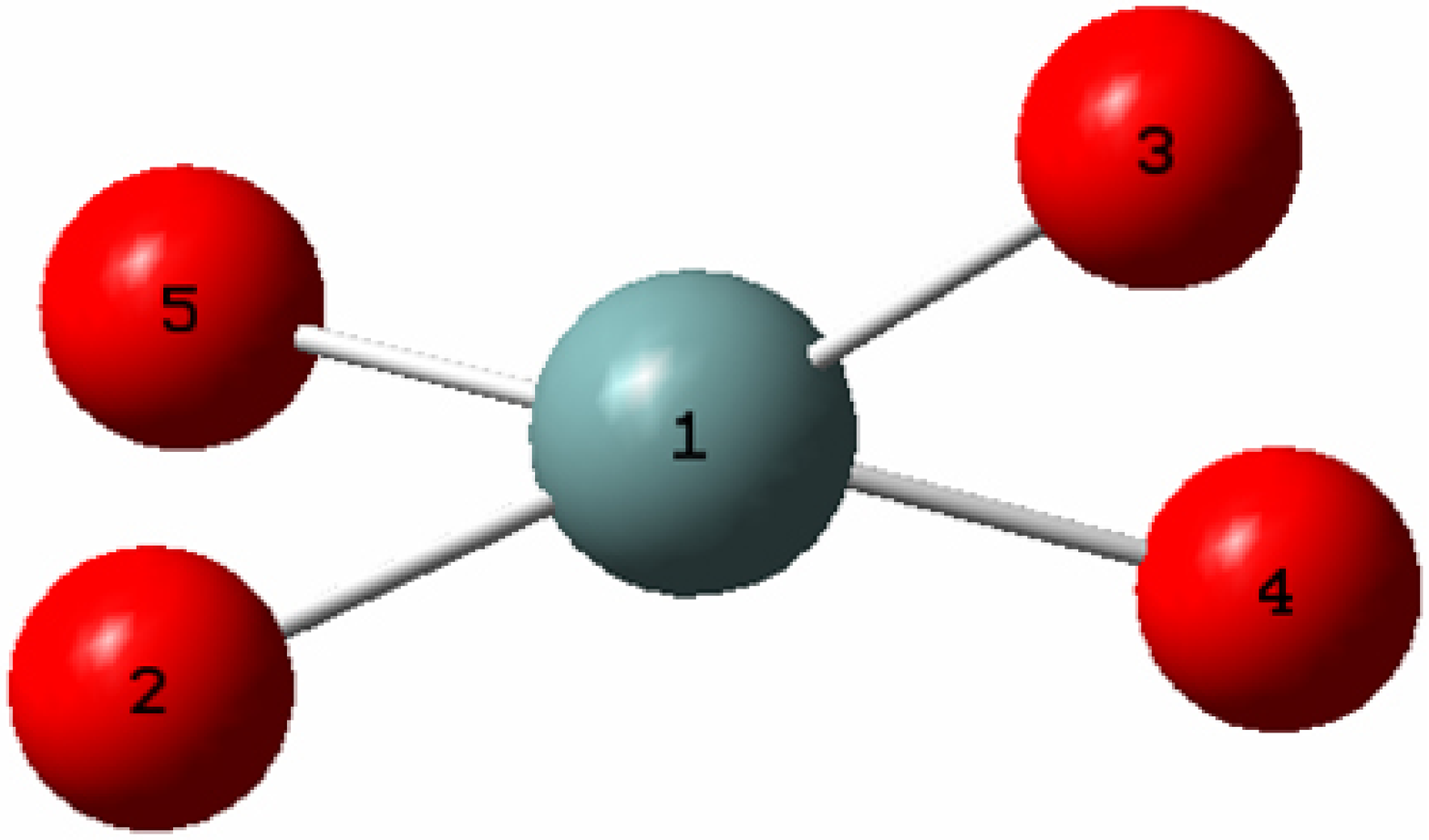}
\includegraphics[width=0.4\columnwidth]{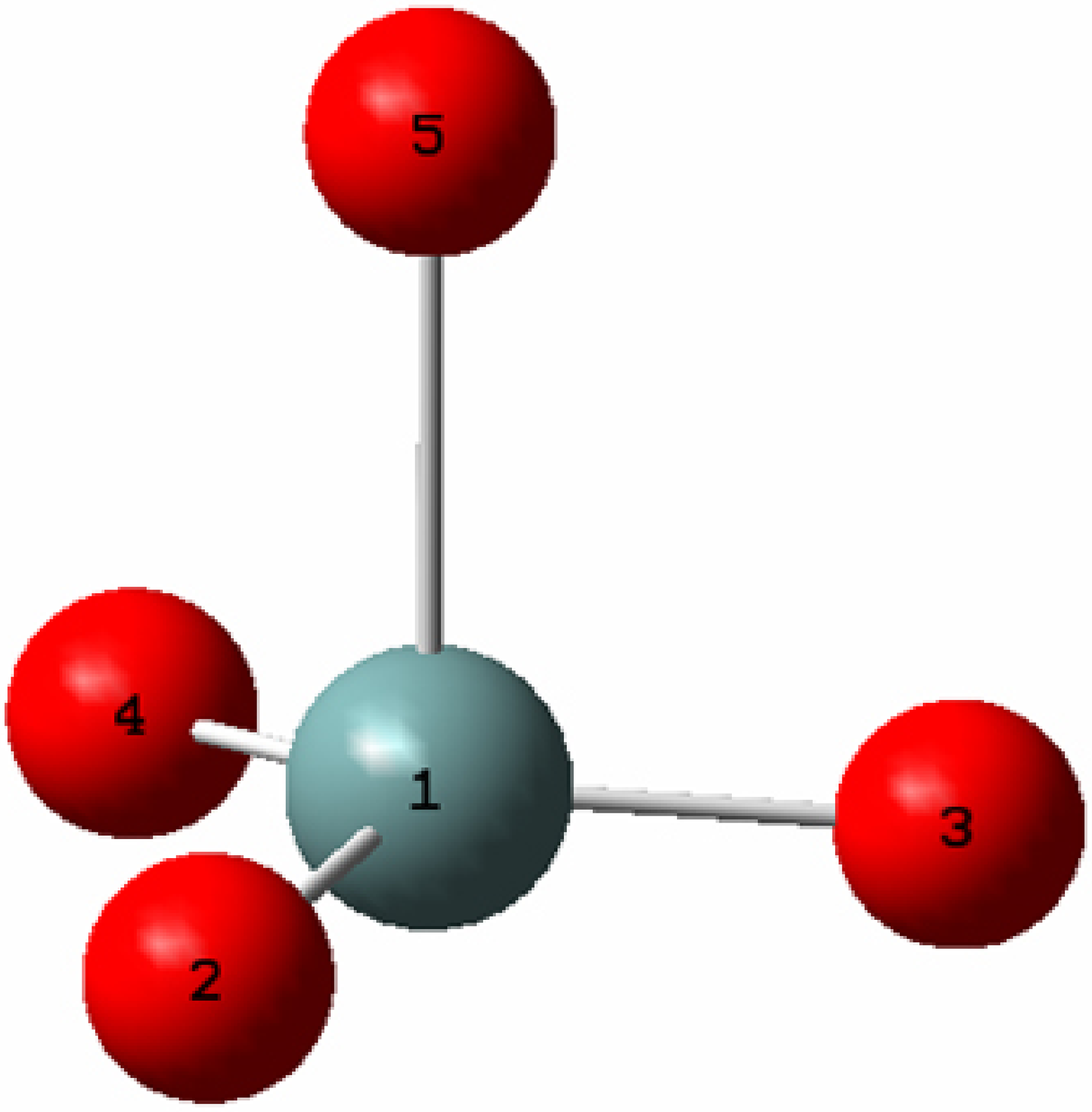}
\caption{(Color online) Low-lying isomers predicted theoretically for neutral
GeO$_4$ and cationic GeO$_4^+$, the total ground-state energies being
$E=-2375.9477$ and $E=-2375.4176$~a.u., respectively. Bond lengths and angles are
given in Tab.~\ref{tab:GeO4p}.}
\label{fig:GeO4p}
\end{figure}

\begin{table}[t]
\centering
\begin{tabular}{cccc}
\hline
 & distance (\AA) & & angle ($^\circ$) \\
\hline
\multicolumn{4}{l}{GeO$_4$} \\
1--2 &    1.770 &   2--1--4 &  	140.31\\
1--3 &    1.770 &   2--1--5 &  	57.32 \\
1--4 &    1.770 &   2--1--3 &  	140.35\\
1--5 &    1.770 &  	    &  		 \\
\hline
\multicolumn{4}{l}{GeO$_4^+$} \\

1--2 &    1.755 &    2--1--4 &   120.58\\
1--3 &    1.755 &    2--1--5 &   95.9  \\
1--4 &    1.755 &    2--1--3 &   118.13\\
1--5 &    2.252 &  	     &  	    \\
\hline
\end{tabular}
\caption{Bond lengths and angles for neutral cluster GeO$_4$ and cation
GeO$_4^+$, as shown in Fig.~\ref{fig:GeO4p}.}
\label{tab:GeO4p}
\end{table}

We show therefore in Fig.~\ref{fig:SiO4p} the proposed structures for these
neutral SiO$_4$ and cationic clusters SiO$_4^+$ from the present
quantum-chemical technique. Bond lengths and angles are correspondingly recorded
in Tab.~\ref{tab:SiO4p}. Analogous results are then shown in
Fig.~\ref{fig:GeO4p} and Tab.~\ref{tab:GeO4p} for neutral GeO$_4$ and cationic
clusters GeO$_4^+$. In the light of the earlier discussion, we note in
particular that now two bond lengths are in evidence, the longer being 2.2~\AA{}
for GeO$_4^+$.

Additionally in Table~\ref{tab:Mulliken} we record our calculations of the
Mulliken charges, which demonstrate the expected shift of electronic charge from
Ge to the O nuclei, due to the high electronegativity of the latter atom.

\begin{table}[t]
\centering
\begin{tabular}{crcrcrcrcr}
\hline
\multicolumn{2}{c}{SiO$_4$} &
\multicolumn{2}{c}{SiO$_4^+$} &
\multicolumn{2}{c}{SiO$_6$} &
\multicolumn{2}{c}{GeO$_4$} &
\multicolumn{2}{c}{GeO$_4^+$} \\
\hline
	Si &	 $2.458$ &		  Si  &    $2.202$  &		Si &     $2.440$ &    
	  Ge &	  $2.416$ &		   Ge  &	$2.348$   \\
	O  &	$-0.615$ &		  O   &   $-0.126$  &		O  &	  $-0.600$ &    
	  O  &	 $-0.604$ &		   O   &   $-0.420$   \\
	O  &	$-0.615$ &		  O   &   $-0.126$  &		O  &	  $-0.600$ &    
	  O  &	 $-0.604$ &		   O   &   $-0.421$   \\
	O  &	$-0.615$ &		  O   &   $-0.474$  &		O  &	  $-0.020$ &    
	  O  &	 $-0.604$ &		   O   &   $-0.424$   \\
	O  &	$-0.615$ &		  O   &   $-0.474$  &		O  &	  $-0.600$ &    
	  O  &	 $-0.604$ &		   O   &   $-0.084$   \\
	   & & &		&		 O	& $-0.600$ & & & & \\
	   & & &		&		 O	& $-0.020$ & & & & \\
\hline
\end{tabular}
\caption{Mulliken charges in free space neutral and cationic clusters.}
\label{tab:Mulliken}
\end{table}

We conclude this section by recording for both GeO$_4$ and GeO$_4^+$ the
one-electron self-consistent field eigenvalues for these neutral and also
charged clusters (Tab.~\ref{tab:eigenvalues}). As with the bond lengths, even
though we are not dealing with stable gas phase molecules like SiH$_4$ and its
cation, there are parallels with the Jahn-Teller removal of degeneracy.

\begin{table}[t]
\centering
\begin{tabular}{|rr|rrrr|}
\hline
\multicolumn{2}{|c|}{GeO$_4$} & 
\multicolumn{4}{c|}{GeO$_4^+$} \\
\hline
$-405.39349$ & $-1.79674$  & $-405.63174$  &  $-20.88425$ &  $-2.03094$ &  $-0.95041$ \\
$-52.29426 $ & $-1.79585$  & $-405.63137$  &  $-20.88342$ &  $-2.03063$ &  $-0.93999$ \\
$-46.38488 $ & $-1.44388$  & $-52.53368 $  &  $-7.58331 $ &  $-2.02971$ &  $-0.93207$ \\
$-46.38488 $ & $-1.43636$  & $-52.53365 $  &  $-7.58274 $ &  $-2.02948$ &  $-0.90334$ \\
$-46.38119 $ & $-1.2333 $  & $-46.62474 $  &  $-5.56251 $ &  $-1.6546 $ &  $-0.89927$ \\
$-20.68764 $ & $-1.23317$  & $-46.62464 $  &  $-5.56214 $ &  $-1.64433$ &  $-0.86177$ \\
$-20.68758 $ & $-0.7747 $  & $-46.62181 $  &  $-5.55912 $ &  $-1.63354$ &  $-0.84468$ \\
$-20.6875  $ & $-0.67349$  & $-46.62164 $  &  $-5.55908 $ &  $-1.57882$ &  $-0.83861$ \\
$-20.68743 $ & $-0.65732$  & $-46.62148 $  &  $-5.55805 $ &  $-1.57679$ &  $-0.83284$ \\
$-7.34623  $ & $-0.65728$  & $-46.62147 $  &  $-5.5578  $ &  $-1.48633$ &  $-0.83033$ \\
$-5.32489  $ & $-0.56935$  & $-21.00925 $  &  $-2.03818 $ &  $-1.47287$ &  $-0.81215$ \\
$-5.32489  $ & $-0.56444$  & $-21.00894 $  &  $-2.03762 $ &  $-1.4617 $ &  $-0.80145$ \\
$-5.32224  $ & $-0.55312$  & $-20.92582 $  &  $-2.03492 $ &  $-0.9934 $ &  $-0.78798$ \\
$-1.80179  $ & $-0.55302$  & $-20.92566 $  &  $-2.03362 $ &  $-0.98829$ &  $-0.78047$ \\
$-1.80075  $ & $-0.53758$  & $-20.9248  $  &  $-2.03347 $ &  $-0.98721$ &  $-0.77532$ \\
$-1.80075  $ & $-0.52855$  & $-20.88442 $  &  $-2.03286 $ &  $-0.9819 $ &  \\
\hline				
\end{tabular}			
\caption{One-electron eigenvalues of both GeO$_4$ and GeO$_4^+$ (atomic units).}
\label{tab:eigenvalues}
\end{table}

\section{Summary}

Our major conclusion from the present theoretical study concerns the predicted
low-lying isomers for the neutral free-space clusters SiO$_n$ ($1<n\leq 6$), and
in briefer fashion for the corresponding Ge clusters. As stressed in
Sec.~\ref{sec:introduction}, the immediate motivation was afforded by the recent
theoretical work of Saito and Ono \cite{Saito:11} on three different crystal
structures of SiO$_2$ under pressure (see also our own early study on
diffraction from vitreous silica in \cite{Stenhouse:76}). It is to be hoped that
our present results will stimulate experimental studies on the free-space
clusters SiO$_n$ for small $n$, and possibly also on the comparison between the
electron density distribution in crystalline SiO$_2$ under pressure with
vitreous SiO$_2$, also under compression.

\section*{Acknowledgements}

The authors thank Dr Saito for correspondence and for providing them with
unpublished supplementary information to Ref.~\cite{Saito:11}. They also thank
Dr Tse for providing them Ref.~\cite{Tse:11} prior to publication, and for
useful discussions. NHM is partially supported by the University of Antwerp (UA)
through the BOF-NOI, and wishes to thank Professors D. Lamoen and C. Van~Alsenoy
for thereby making possible his continuing affiliation with UA. NHM also thanks
Professors R. Pucci and G. G. N. Angilella for generous hospitality during his
stay in Catania.

\bibliographystyle{mprsty}
\bibliography{a,b,c,d,e,f,g,h,i,j,k,l,m,n,o,p,q,r,s,t,u,v,w,x,y,z,zzproceedings,Angilella}

\end{document}